\documentclass[acmtog]{acmart}
\acmSubmissionID{1161}
\usepackage{booktabs} 
\usepackage{hyperref}

\copyrightyear{2026}
\acmYear{2026}
\setcopyright{cc}
\setcctype{by-nc-nd}
\acmConference[SA Conference Papers '26]{SIGGRAPH Asia 2026 Conference Papers}{December 01--04, 2026}{Kuala Lumpur, Malaysia}
\acmBooktitle{SIGGRAPH Asia 2026 Conference Papers (SA Conference Papers '26), December 01--04, 2026, Kuala Lumpur, Malaysia}
\acmDOI{10.1145/3829340.3842175}
\acmISBN{979-8-4007-2842-6/2026/12}

\setcitestyle{square}

\usepackage{natbib}

\usepackage{graphics}
\usepackage{multirow}

\usepackage{pifont}
\newcommand{\topscore}[1]{\textcolor{blue}{\textbf{#1}}} 
\newcommand{\udl}{\underline} 

\usepackage{graphicx}

\usepackage{float}

\usepackage{atbegshi}
\newcommand{\placetextbox}[3]{
  \setbox0=\hbox{#3}
  \AtBeginShipoutNext{\AtBeginShipoutUpperLeft{%
    \put(\dimexpr#1\paperwidth\relax,-\dimexpr#2\paperheight\relax)
    {\vtop{{\null}\makebox[0pt][c]{#3}}}%
  }}%
}

\usepackage[capitalise,nameinlink, noabbrev]{cleveref}
\newcommand{\ua}{\uparrow}
\newcommand{\da}{\downarrow}

\newcommand{\myhref}[3][blue]{\href{#2}{\color{#1}{#3}}}
\newcommand{\imagecredits}[1]{\textcolor{darkgray}{\phantom{a} \hfill \small{Image credits: #1}}}

\begin{document}

\title{Recurrent Dynamic Range Extension}

\author{Sebastian Dille}
\affiliation{%
 \institution{Computational Photography Lab., Simon Fraser University}
 \country{Canada}}

\author{Keru Fu}
\affiliation{%
 \institution{Computational Photography Lab., Simon Fraser University}
 \country{Canada}}

\author{S. Mahdi H. Miangoleh}
\affiliation{%
  \institution{Computational Photography Lab., Simon Fraser University}
  \country{Canada}}

\author{Yağız Aksoy}
\affiliation{%
  \institution{Computational Photography Lab., Simon Fraser University}
  \country{Canada}}

\renewcommand{\shortauthors}{Dille et al.}

\begin{abstract}
We present an approach to progressively extend the highlights of an image. Instead of reconstructing the full dynamic range of a complex scene directly, we learn a simpler task first: We extend the dynamic range of an input image by a single exposure value. Once this is mastered, we retrieve the full HDR image for the scene by executing our network recurrently, progressively increasing the dynamic range of the input. Our formulation is agnostic to the input dynamic range and targets a bounded output domain. This enables us to use widely available RAW images for the reconstruction task and adapt adversarial losses to construct realistic images. By incorporating Memory Replay for backpropagation, we can train our network recurrently over multiple inference stages and reduce reconstruction errors. As a consequence, our system reconstructs challenging long-tailed HDR scenes robustly and shows powerful recovery of bright light sources and highlights.
\end{abstract}


\begin{teaserfigure}
  \centering
  \includegraphics[width=\linewidth]{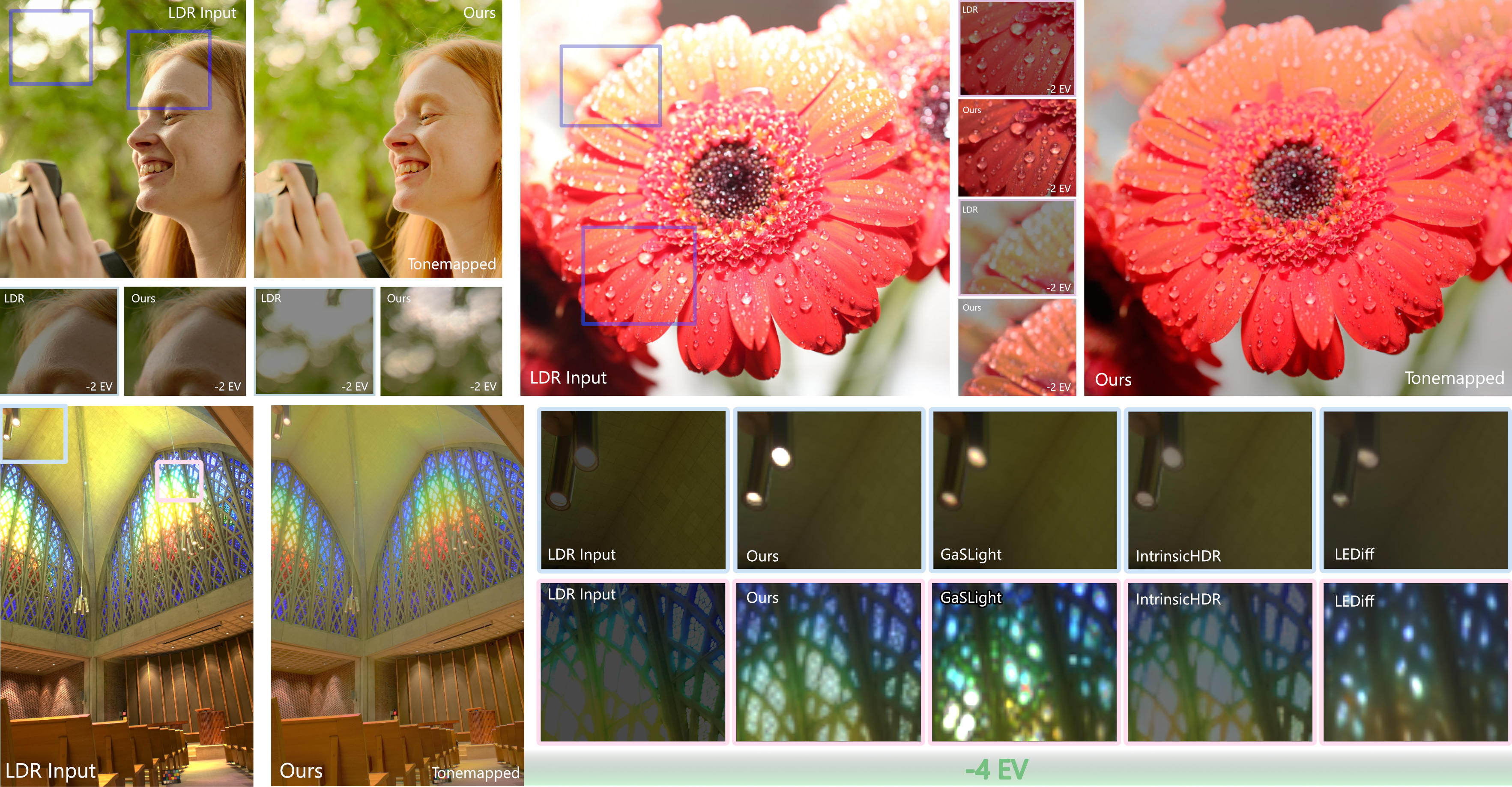}
  \caption{
  We present a recurrent highlight reconstruction approach that progressively expands the dynamic range of a single LDR image. By iteratively extending the range, our method recovers missing information in severely clipped regions, including specular highlights, strong light sources, and reflective surfaces. Trained on RAW data in addition to HDR ground-truth, our model generalizes well across diverse scenes and outperforms prior methods in restoring high-intensity details while maintaining consistent structure and color fidelity under large exposure changes. Results are tone-mapped for visualization. 
  \phantom{empty} \imagecredits{
  \myhref[darkgray]{https://www.pexels.com/photo/a-woman-using-her-camera-8764382/}{@$\text{Darina Belonogova}$},  \citet{fairchild2007hdr}
  }
  }
  \label{fig:teaser}
\end{teaserfigure}
\maketitle

\placetextbox{0.14}{0.03}{\includegraphics[width=4cm]{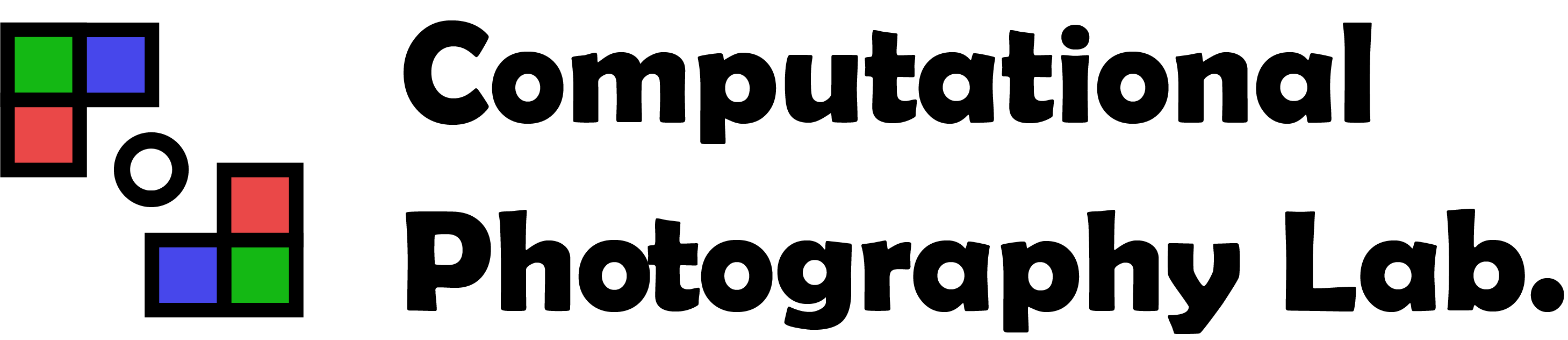}}
\placetextbox{0.8}{0.03}{Find the project web page here:}
\placetextbox{0.8}{0.045}{\textcolor{purple}{\url{https://yaksoy.github.io/RecurrentHDR/}}}

\begin{figure*}
  \includegraphics[width=\linewidth]{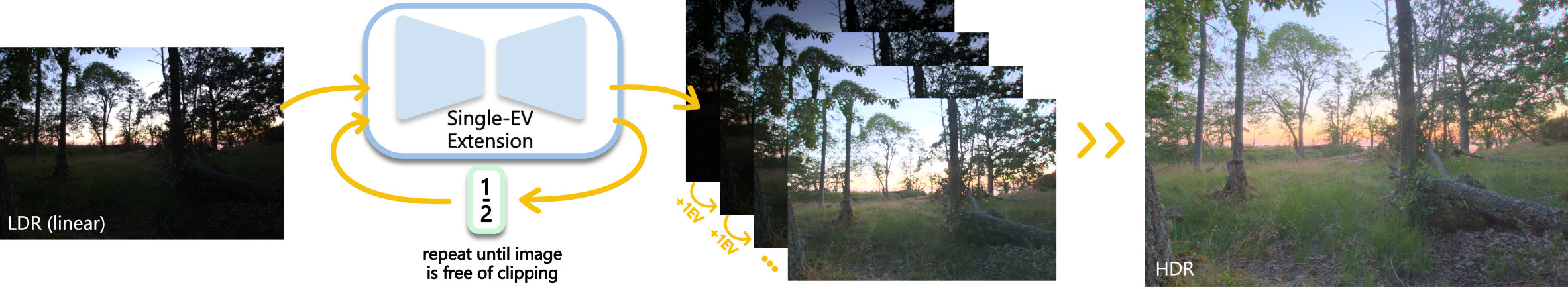}
  \caption{
    Starting from a single LDR input, we iteratively apply our single-step dynamic range extension network. After each iteration, the predicted image serves as input again to further expand the dynamic range. This recurrent process progressively restores clipped regions and is repeated until the image is free of clipping and the radiance range of the scene is recovered. Results are tone-mapped for visualization.
    \imagecredits{SI-HDR benchmark~\cite{hanjiComparisonSingleImage2022}}
  }
  \label{fig:pipeline_fwd}
\end{figure*}
\section{Introduction}
Natural scenes often exhibit an extremely wide dynamic range. Radiance values in reflective regions, specular highlights, or visible light sources can be several orders of magnitude larger than the rest of the scene, leading to unbounded long-tailed intensity distributions.
Modern imaging sensors are incapable of capturing the full dynamic range of natural scenes in a single photo. 
In common photographs, when the radiance in a scene exceeds the sensor capacity, high-intensity regions are clipped, producing low-dynamic-range (LDR) images that lack information in bright areas.

Data-driven approaches provide a promising solution for inferring the missing content from casual observations. However, this relies on large and diverse training data to build a strong prior of the radiance in natural scenes. The necessary high dynamic range (HDR) ground truth is typically obtained through multi-exposure bracketing, which requires static scenes, precise alignment, and tripod-mounted cameras. This capturing process is time-consuming and error-prone, significantly limiting the scale and diversity of available real-world HDR data. Thus, existing datasets remain scarce and lack the rich diversity needed to train generalizable models. 

At the same time, the long-tailed radiance distribution of HDR scenes poses a challenge for common image-based losses, such as L1 or L2. The high luminance of specular highlights or active light sources can lead to large prediction errors in early training and exploding gradients during backpropagation. The corresponding large weight updates destabilize the training process, since the optimization is likely to oscillate or diverge. Consequently, directly regressing the HDR image requires tone mapping to compress the luminance and stabilize the training. While highlight compression is principally similar to the behavior of the human visual system, it is prone to reducing physical accuracy: The compression hinders local differentiation within the highlight regions, and small inaccuracies in this compressed space translate into large errors in the linear radiance values. For these reasons, recovering the clipped highlights from a single LDR input is a challenging problem.

In this paper, we propose learning a much simpler task first: Given a linear clipped input image, we train a neural network to extend its highlights by a single exposure value (EV). The clipped input $\in [0,1]$ becomes an extended image $\in [0,2]$, independent of the nominal exposure value of the input and its initial dynamic range. The network learns to restore the clipped values of saturated pixels, while bright highlights remain clipped if they exceed the new bounds. This single-EV extension is exposure invariant and constrained by design. Since the output is bounded $\in [0,2]$, we can train our network in \emph{linear space} without risking exploding gradients. It allows us to omit any compression of the target domain and employ common perceptual losses to enhance realism and preserve visual fidelity. Likewise, extending a clipped linear image to a slightly extended dynamic range can be trained on diverse RAW photo collections. While RAW images do not capture the full dynamic range of a scene, they preserve significantly more radiometric information than standard 8-bit LDR images. We can hence augment our training data by leveraging the exposure-invariant formulation and derive multiple training pairs from a single re-exposed RAW photo. Together, this provides our extension network with a rich and diverse training distribution to enable generalization to in-the-wild photographs.

Crucially, we can move beyond the bounds of our single-EV extension by applying our network recurrently over multiple inference steps. This enables us to progressively recover the full long-tailed distribution of HDR highlights by using the output of one step as input to the next. For this purpose, we implement a recurrent training scheme with supervision from real HDR images. Our network thus expands its learned prior from RAW-based training with the continuous highlight distribution from HDR data. It learns to extend the first and the $n$-th input of the recurrent execution adaptively, while compensating for prediction errors in previous inference steps.

Our contributions are as follows: 

We present a novel exposure-invariant recurrent highlight extension that effectively tackles the main challenges for single-image highlight reconstruction. The unbounded scene luminance is broken down into clearly constrained, separate chunks, which can be trained on RAW images to improve generalization. We further incorporate a recurrent training scheme on HDR images to enable backpropagation for our full ResNet-like architecture over multiple inference steps. With our recurrent training, we can combine the separate steps into a consistent reconstruction of the long-tailed luminance. We show examples in Figure~\ref{fig:teaser}. Our method faithfully recovers the skin tone of the woman as well as the bokeh pattern in the background. On the flower, it recovers the color visible through the raindrops while preserving the bright reflections. For the church, it predicts accurate spotlights and a bright sunlit window. Our resulting luminance for the spotlights thereby still exceeds the displayable dynamic range, even with an exposure correction of 4 EV. As we will further demonstrate in Section~\ref{sec:exp}, this design enables accurate and generalizable highlight reconstruction in high resolution for real-world, in-the-wild images. We show an overview of our pipeline in Section~\ref{fig:pipeline_fwd}, and describe the inference and training details in Section~\ref{sec:multi_step}.

\section{Related Work}
\textbf{Inverse tone mapping.}\quad
Reconstructing clipped areas is a long-standing research problem. Earlier approaches \cite{banterle2006inverse,rempel2007ldr2hdr,banterle2008expanding,didyk2008enhancement,masia2009evaluation,masood2009automatic,abebe2018towards} aim to create HDR intensities in the highlights with local-adaptive~\cite{banterle2006inverse,kovaleski2009high,rempel2007ldr2hdr,didyk2008enhancement,banterle2008expanding} or global brightness scaling~\cite{masia2009evaluation}, heuristically determining the target intensity. Some additionally fill in color via cross-channel correlation~\cite{masood2009automatic,abebe2018towards}. While enhancing the LDR input, these methods do not reproduce details and struggle to generalize to natural scenes.

\noindent\textbf{LDR-to-HDR regression.}\quad
With data-driven methods, a straightforward approach to highlight reconstruction is to predict the full dynamic range of the scene directly, including enhanced shadow areas. Such methods operate either in a single step~\cite{eilertsenHDRImageReconstruction2017, marnerides2018expandnet,santosSingleImageHDR2020, yu2021luminance, chen2021hdrunet,Marnerides2021DeepHH, guo2022lhdr,zou2023rawhdr} or by breaking down the reconstruction into different sub-tasks~\cite{liu2020single,zhang2021deep,dilleIntrinsicHDR}. In both settings, the unbounded range of the highlights poses the main challenge for the optimization. Existing approaches consequently operate in log-space~\cite{eilertsenHDRImageReconstruction2017, santosSingleImageHDR2020, liu2020single, yu2021luminance} or apply tone-mapping~\cite{chen2021hdrunet,marnerides2018expandnet,Marnerides2021DeepHH, guo2022lhdr, dilleIntrinsicHDR} to compress the dynamic range. Additionally, direct regression relies exclusively on HDR data for training to learn the typical long-tailed distribution. With the limited availability of HDR datasets and the additional compression during training, the methods do not generalize well to images in the wild and tend to saturate early for specular highlights or visible light sources. 
Our method focuses specifically on the challenging highlights. It operates within a bounded linear domain, avoiding outliers in the loss function and exploding gradients. Our network faithfully learns a physically accurate HDR distribution without requiring any nonlinearities for supervision. Most importantly, we combine the diversity of RAW image collections and the full radiance information from HDR datasets to train a model that generalizes to everyday photographs through our 2-step training scheme.

\noindent\textbf{Exposure-stack reconstruction.}\quad
A different direction is the reconstruction of exposure stacks from a single image~\cite{endoDeepReverseTone2017,lee2018deep,leeDeepRecursiveHDRI2018,lee2020learning,jo2021deep,kim2021end,le2023single,cevr_2023, gaslight}, which can later be merged into an HDR image using established multi-exposure merging~\cite{mann1995being,debevec1997recovering}. Closely related to our approach, these methods create differently exposed \emph{LDR} images from a given input. This circumvents the challenge of the unbounded target domain and enables the use of common losses without additional compression. 
Early approaches~\cite{endoDeepReverseTone2017,lee2018deep} are trained to predict a fixed set of output exposures simultaneously. They are restricted to a defined dynamic range and require architectural changes and retraining on HDR data to adapt to other scenes. Succeeding works estimate the exposures individually, either by conditioning the estimation on a given target exposure~\cite{cevr_2023,kim2021end} or only differentiating between increasing and decreasing the exposure in a recursive inference setup~\cite{leeDeepRecursiveHDRI2018,jo2021deep, gaslight}. The iterative execution of the same network, however, is prone to accumulated estimation errors and severe reconstruction artifacts. 

Critically, these methods do not \emph{extend} the dynamic range but rather create LDR outputs with \emph{shifted} upper and lower bounds. Being supervised on stacked LDR images, the networks learn not only to fill in pixels but also to \emph{remove} valid content from the input to create new under- or over-exposed areas. This behavior is encouraged by cycle-consistent losses~\cite{lee2020learning,cevr_2023} but increases the risk of reconstruction errors by removing valuable information. As a result, the methods perform worse on HDR benchmarks such as SI-HDR~\cite{hanjiComparisonSingleImage2022} when compared to direct-regression approaches. 
In contrast, our approach progressively \emph{extends} the dynamic range, reducing the ambiguity at each step without removing valuable information. Our exposure-invariant formulation enables our network to adapt the dynamic range in a \emph{scene-specific} manner, instead of using ensembles specialized for individual steps~\cite{lee2018deep} or conditioning on the arbitrary range of a target HDR image~\cite{leeDeepRecursiveHDRI2018}. Our \emph{recurrent training scheme} allows us to adaptively reconstruct the dynamic range of the highlights without restrictions from the architecture.

\noindent\textbf{Generative approaches.}\quad
Recently, approaches have been introduced that adapt the multi-exposure generation practice to fine-tune generative diffusion models~\cite{bemana2025bracket, wang2025lediff, gaslight}. They benefit from their strong pretrained image prior and apply the constrained exposure-estimation framework to directly generate HDR images. For HDR reconstruction, these approaches fill in clipped areas more effectively, where non-generative methods lack details. At the same time, their prior is learned on \emph{non-linear} data and shows a distribution mismatch to the HDR domain. The models are prone to modifying the image content and creating artifacts. The generative capabilities come at the cost of large networks and a minute-long inference process, while the image resolution is limited.
Our recurrent extension is independent of the architecture. Using fully-convolutional networks and training in the \emph{linear} domain, it shows strong generalization to diverse scenarios under challenging settings. We can operate at high resolutions, reconstructing the highlights of a 4k image in half a second, while remaining faithful to the original image.

\noindent\textbf{Recurrent training in computational photography.}\quad
Recurrent approaches are typically used to model time series in low-dimensional regimes like NLP~\cite{hochreiter1997long,cho2014properties,shiva2018survey}. Their training is often limited due to the high computational burden of backpropagating gradients over multiple inference stages. For computational photography applications, existing approaches remain scarce~\cite{shi2020calibrcnn,cao2023recurrent,wang2017deepvo,teed2020raft,hidalgo2020learning} and rely on low resolutions and shallow recurrent modules to reduce complexity. The same applies to existing HDR reconstruction approaches~\cite{kim2021end}, limiting their applicability for professional image editing.
In our training, we adapt the Memory Replay strategy~\cite{memformer} from NLP literature to the image domain, reducing the compute for recurrent backpropagation significantly. It allows us to train our full ResNet-based network recurrently without sacrificing resolution or network capacity.

\section{Recurrent Dynamic Range Extension}
\label{sec:multi_step}
\begin{figure}[t]
  \includegraphics[width=\linewidth]{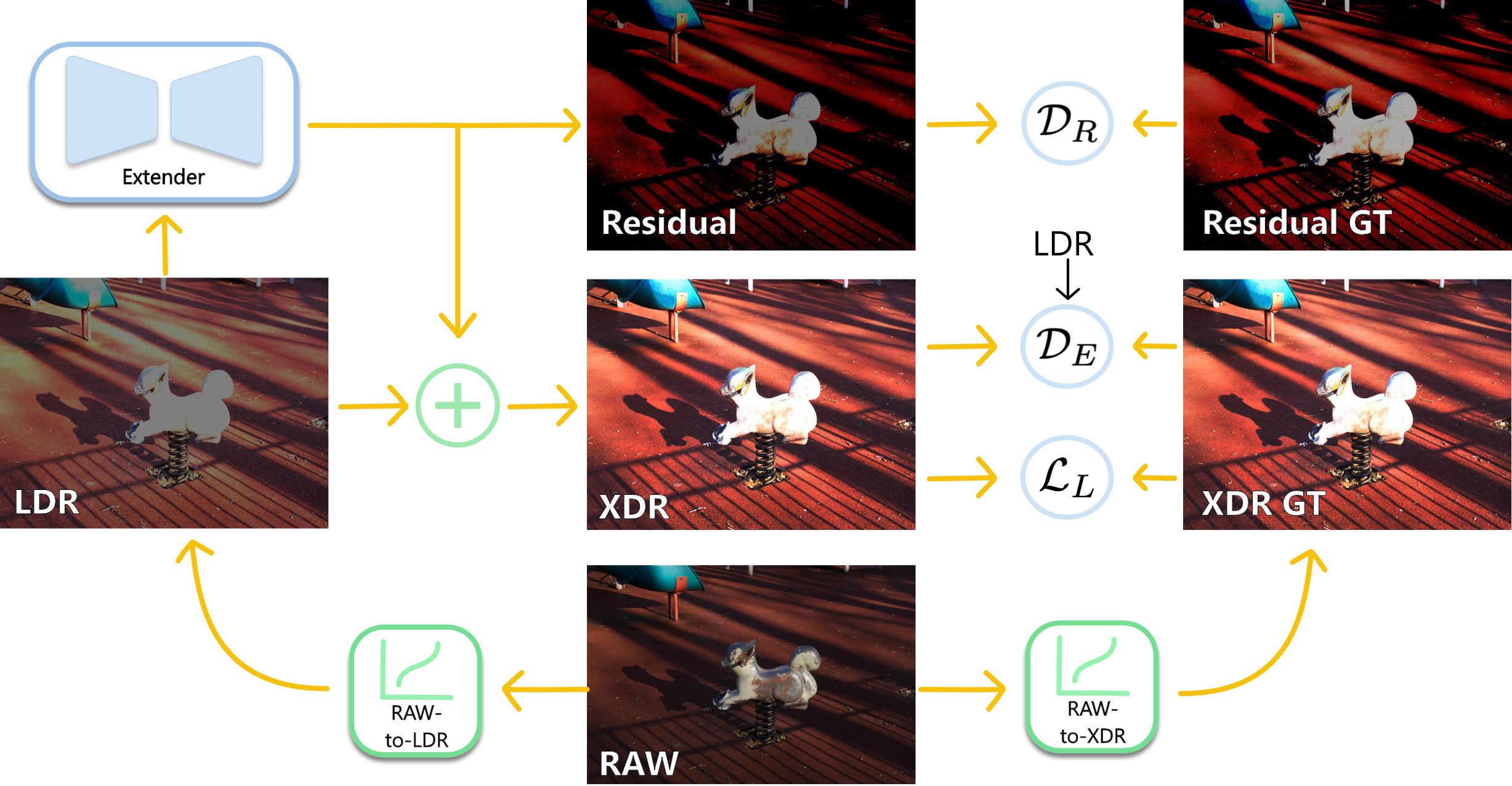}
  \caption{
     Training the single-step dynamic range extension. From a given RAW photo, we create an LDR input and an extended target. Our network then predicts a residual image that is added to the LDR input to reproduce the extended image. We supervise both the residual and the extended output using a dual-discriminator adversarial loss to ensure realism.
     \newline \imagecredits{\citet{fivek}}     
     }
  \label{fig:single_stage}
\end{figure}

We formulate the reconstruction of clipped highlights as an iterative process of multiple bounded extensions of the input dynamic range:
We start with a clipped input image $I_L \in[0,1]$ in linear RGB space. At each inference step, our core extension model is tasked to increase the luminance range of the input by exactly one EV, yielding an extended image $I_E \in[0,2]$. Instead of directly estimating the extended image, we design our extension network to predict a residual image $I_R$, and compute the extended result by adding this residual to the input, i.e., $I_L + I_R = I_E$. We visualize this process in~\cref{fig:single_stage}. Our formulation provides a straightforward mechanism for the network to recover information clipped in $I_L$, rather than redundantly recreating content that is already present in the input. By only estimating the missing information, the residual formulation naturally preserves other image regions and also keeps the network output bounded within $[0,1]$.

In our recurrent pipeline, we repeat this process multiple times, dividing the output by two and providing it as input to the same network, until the result is free of clipping and the full luminance range is restored. Lastly, we revert the repeated normalization and use a soft mask $M$ to linearly blend our result over the input image to compensate for any reproduction artifacts in the shadows, with $M=max(0,I_{L}-0.5)/0.5$. This yields the final HDR estimate $\hat{I}_H$ of our method. Note that at inference, the number of network iterations is not fixed and fully depends on the content of the scene. We stop when our network does not leave clipped pixels in its result, i.e., when the maximum value of the output does not reach our empirical threshold of $1.8$.

\subsection{Learning Single-EV Extension}
\label{sec:single_step:loss}
Our extension mechanism relies on a robust prior over diverse natural images. For this purpose, we generate input and target pairs $I_L$ and $I_E$ that are one EV apart directly from RAW data collections. RAW images represent the linear irradiance captured by the camera without any display-referred compression. With the common 14-bit signal processing of modern sensors, they offer a wider dynamic range than the final LDR photo and yield suitable training targets for our extension after color reproduction. At the same time, they do not require the controlled setup necessary for exposure stacks, but can be readily captured and stored. Publicly available RAW collections~\cite{raise, fivek, SID, hdrplus, ppr, canonraw, nikonraw} consequently exceed HDR datasets drastically in both size and diversity.

Our constrained extension steps allow us to supervise the reconstruction in linear space. Therefore, we directly follow the adversarial scheme proposed by \citet{huang2024gan} for supervision to encourage realistic results.  We apply LPIPS \cite{zhang2018unreasonable} as a pixel-wise loss ($\mathcal{L}_{LPIPS}$) and use two distinct discriminators $\mathcal{D}_E$ and $\mathcal{D}_R$ to ensure the realism and plausibility of $I_E$ and $I_R$, respectively. We also condition $\mathcal{D}_E$ on the input image $I_L$. Our combined training loss can be summarized as:
\begin{equation}
    \label{eq:train_loss}
    \mathcal{L} = \mathcal{L}_{LPIPS} + \mathcal{D}_E + \mathcal{D}_R.
\end{equation}
 The combination of our two discriminators with their complementing properties is essential to avoid degenerate solutions. Recall that the extended image can still contain clipping from bright highlights exceeding the extended range, as it only increases dynamic range by one EV. Hence, the output characteristics are similar to those of $I_L$, leaving several options for the extension network to fool the discriminator. Adding $I_L$ as a conditional input to the $\mathcal{D}_E$ serves as an important clue to identify a realistic extension. It prevents the network from producing a trivial or all-zero residual that would result in a simple reproduction of $I_L$. 

\begin{figure*}
  \includegraphics[width=\linewidth]{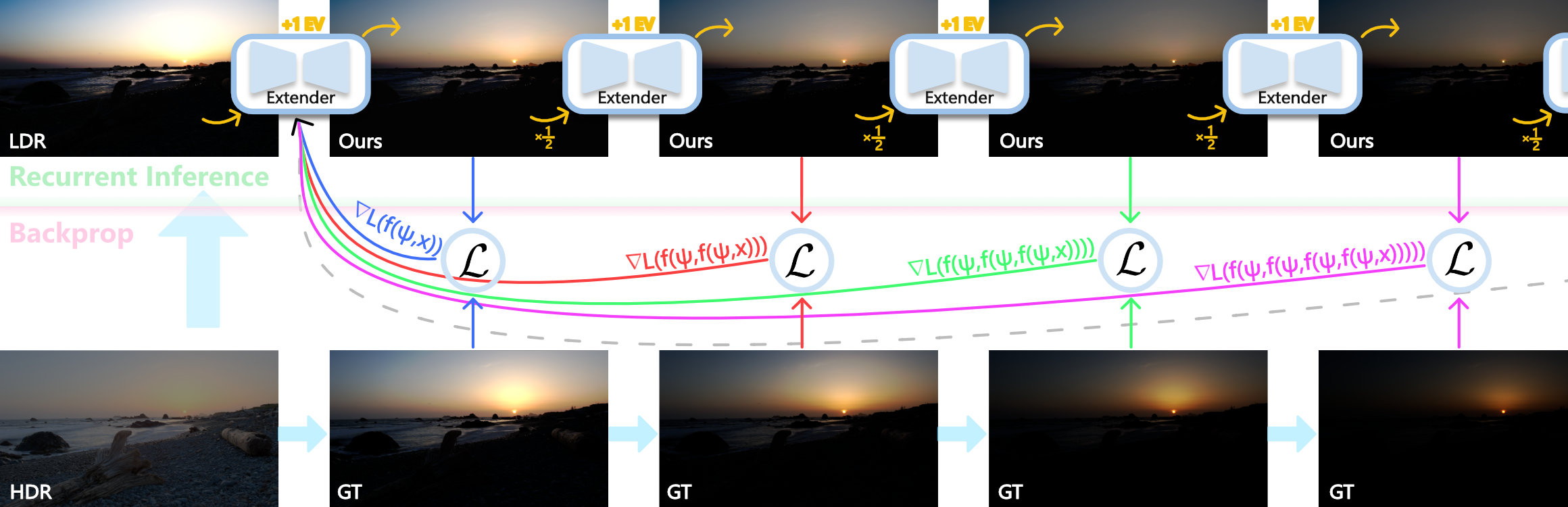}
  \caption{
     Our recurrent training pipeline. We create a linear LDR input and targets for $N$ steps from an HDR image. Next, we iteratively perform inference using the prediction of each step as input to the next. We then backpropagate the reconstruction error through the entire inference chain, taking the contribution of all steps into account.
    \imagecredits{\citet{fairchild2007hdr}}
  }
  \label{fig:multi_stage}
\end{figure*}

On the other hand, cases such as fully white estimates or duplicating $I_L$ in $I_R$ yield a simple reproduction of the input with shifted or scaled brightness. These results are degenerate because the dynamic range remains unchanged, but they resemble real images and appear different from the original $I_L$. $\mathcal{D}_R$ imposes a constraint on the distribution of the generated residuals and mitigates this effect. At the same time, a single discriminator for the residuals cannot guarantee that the extended image $I_E$ itself looks realistic. It also will not penalize the extension network for modifying pixels not affected by clipping. Using this design, we train our dynamic range extension model with a batch size of 20 for 400 epochs. We provide additional details for our data processing and our losses and an ablation for our dual-discriminator setup in the supplementary material.

\subsection{Learning Accurate Highlight Distributions}
\label{sec:multi_step:arch}
Our single-exposure extension already provides the network with a prior over diverse RAW images. To successfully reconstruct true HDR images, it additionally has to learn the realistic luminance of highlights and light sources over a wide dynamic range, and to correct its previous estimation errors in later stages. For this purpose, we integrate the recurrent inference as a second stage into its training process. 

We show an overview of the recurrent training strategy in Figure~\ref{fig:multi_stage}. During training, we repeatedly apply our network for $N$ inference steps, where the normalized output of each step serves as the input to the next. This produces a sequence of intermediate estimates $\{I_E^n\}^N_{n=1}$ defined as $I_E^n = f(\psi, I_E^{n-1})$ where $f(.)$ denotes the network with parameters $\psi$. 
Each intermediate prediction $I_E^n$ is supervised using its corresponding ground truth derived from HDR data. During training, gradients from the loss at each step propagate through the entire chain of iterative network executions. While the result for the first step depends on the network itself (i.e. $f(\psi, I_L)$), the second step becomes $f (\psi,f (\psi, I_L))$, the third $f(\psi, f(\psi( f (\psi,I_L)))$, and so on. By backpropagating through the same network parameters, weight updates also reflect how errors introduced at earlier steps influence subsequent extensions. This allows the model to learn to reduce mistakes made in earlier iterations by propagating corrective gradients back to the beginning of the chain.

This iterative gradient computation leads to a heavy computational burden, which exceeds the memory of most common GPUs. Because of this limitation, previous methods omit backpropagating through recurrent inference, leading to subpar results where prediction errors accumulate over iterations. We efficiently compute the complex recurrent gradients by adapting the Memory Replay strategy from NLP literature~\cite{memformer}, and successfully integrate the recurrent inference in our training. Different from inference, the number of recurrent steps $N$ in training does not depend on the image content. Starting from our single-EV pretraining with $N=1$ and applying the same losses, we progressively increase $N$ by 1 and continue training with the fixed step count. We repeat this process three times to a maximum of $N=4$. After each increment, we train for 100 epochs with the new $N$ to allow our model to adjust. For $N>4$, the number of valid pixels in our training data is insufficient to yield meaningful supervision signals.

During the recurrent training, it is crucial to provide accurate supervision over the wider dynamic range. We construct our training data this time from HDR images that represent specular highlights and light sources well. Our network thus adapts to the long-tailed distribution of HDR images and simultaneously maintains the prior from single-stage training. We use HDRT~\cite{peng2025hdrt}, S2R-HDR~\cite{wang2025s2r}, and HDR-Real~\cite{liu2020single} for our recurrent training scheme. We provide the details for our HDR data processing and the Memory Replay in the supplementary material.

\subsection{Network Architecture And Training Details}
\label{sec:architecture}
Our dynamic range extension model is built upon the fully convolutional MiDaS 2.1 architecture~\cite{midas} with a ResNext101-backbone. It is well-established for high-resolution imaging tasks and has been applied for HDR reconstruction previously \cite{dilleIntrinsicHDR}. For both discriminators, we use the architecture from PatchGAN \cite{patchgan} with 3 layers each. We use AdamW \cite{adamw} as an optimizer and a Cosine Annealing Scheduler \cite{loshchilov2016sgdr} with $T_{max}=10$. The learning rate at the start of training is $10^{-4}$ for the generator and $10^{-5}$ for both discriminators. 

\begin{table*}[t]
\centering
\caption{Quantitative results against state-of-the-art on the SI-HDR~\cite{hanjiComparisonSingleImage2022} and BtP-HDR~\cite{bolduc2023beyond} benchmarks. We compute PSNR, VSI, and P-H in the perceptual uniform PU21-space~\cite{azimi2021pu21}. Grey font color indicates that the BtP-HDR Benchmark was used in training.}

\resizebox{0.90\linewidth}{!}{%
\begin{tabular}{lcccccc|cccccc}
& \multicolumn{6}{c|}{SI-HDR benchmark~\tiny\cite{hanjiComparisonSingleImage2022}} & \multicolumn{6}{c}{BtP-HDR benchmark~\tiny\cite{bolduc2023beyond}} \\
\hline
 Method &   PSNR$\ua$ &    VSI$\ua$ & VDP3$\ua$ &  P-H$\ua$ &  FLIP$\da$ &CVVDP$\ua$
 &PSNR$\ua$ &    VSI$\ua$ & VDP3$\ua$ &  P-H$\ua$ &  FLIP$\da$ &CVVDP$\ua$\\
\hline
IntrinsicHDR~\tiny{\cite{dilleIntrinsicHDR}}                &36.62  &98.27  &\udl{8.96} &32.63  &55.92  &7.46
                                                            &34.86  &97.86  &8.85       &31.48  &59.50  &8.38    \\
HDR-CNN~\tiny{\cite{eilertsenHDRImageReconstruction2017}}   &35.91  &98.17  &8.39       &31.87  &60.98  &6.82   
                                                            &36.18  &\topscore{98.46}   &\udl{9.04} &32.83  &\udl{44.92}    &\topscore{8.88}  \\
ExpandNet~\tiny{\cite{marnerides2018expandnet} }            &36.01  &97.65  &8.67   &32.53  &55.69  &6.95    
                                                            &\udl{36.33}    &98.15  &8.74   &\udl{33.25}    & 67.16 &7.79 \\
Single-HDR~\tiny{\cite{liu2020single}}                      &35.68  &\udl{98.30}    &8.79   &31.34 &\udl{52.40} &\topscore{7.76} 
                                                            &33.74  &97.85  &8.85   &30.00  &70.04  &7.79\\
Mask-HDR~\tiny{\cite{santosSingleImageHDR2020}}             &36.72  &98.22  &8.25   &\udl{33.28} &62.33 &6.75 
                                                            &35.96  &\udl{98.40}    &9.00   &32.88  &\topscore{42.81} &\topscore{8.88}   \\
HDRUNet~\tiny{\cite{chen2021hdrunet}}                       &\udl{36.92}    &98.16  &8.82   &31.58  &85.92  &5.24   
                                                            &35.71  &98.13  &8.90   &32.23  &74.49 &6.76 \\
\hline
DrTMO~\tiny{\cite{endoDeepReverseTone2017}}                 &33.58  &96.73  &8.27   &28.50  &84.58  &5.96   
                                                            &34.83  &97.94  &8.41   &31.39  &71.09  &7.60   \\
Multi-Exp Gen.~\tiny{\cite{le2023single}}                   &35.36  &98.01  &8.64   &31.23  &86.88  &5.08   
                                                            &33.72  &97.83  &8.41   &30.41  &77.43  &6.51  \\
CEVR~\tiny{\cite{cevr_2023}}                                &34.48  &97.31  &8.46   &31.07  &64.99  &6.59   
                                                            &34.32  &97.56  &8.46   &31.0   &63.81  &8.01   \\
DiffSyn~\tiny{\cite{kim2021end}}                            &35.56  &97.66  &8.60   &31.39  &63.22  &6.47     
                                                            &34.15  &97.78  &8.33   &30.51  &65.04  &7.84   \\ 
BracketDiffusion~\tiny{\cite{bemana2025bracket}}            &32.50  &95.52  &7.54   &28.38  &64.62  &6.06     
                                                            &34.69  &97.39  &8.46   &31.85  &53.15  &8.29 \\
LEDiff~\tiny{\cite{wang2025lediff}}                         &32.69  &96.41  &7.89   &28.47  &81.55  &4.63   
                                                            &35.61  &98.04  &8.64   &32.69  &58.37  &7.62 \\
GaSLight~\tiny{\cite{gaslight}}                      &34.92  &97.97  &8.65   &29.95  &55.90  &7.02 
&\textcolor{gray}{37.05}    &\textcolor{gray}{98.25}    &\textcolor{gray}{9.09}      &\textcolor{gray}{33.99}   &\textcolor{gray}{52.41} &\textcolor{gray}{8.86} \\
\hline
OURS    &\topscore{37.94} &\topscore{98.79}     &\topscore{9.10}    &\topscore{35.09}   &\topscore{48.63} &\udl{7.64} 
        &\topscore{36.37} &\topscore{98.46}     &\topscore{9.12}    &\topscore{33.42}   & 49.92 &\udl{8.87} \\
\hline
\end{tabular}
}

\label{tab:all_sihdr}
\end{table*}

\section{Experiments}
\label{sec:exp}

We evaluate the performance of our method by comparing it against a wide set of existing approaches. HDR-CNN~\cite{eilertsenHDRImageReconstruction2017}, ExpandNet~\cite{marnerides2018expandnet}, HDRUnet~\cite{chen2021hdrunet}, and Mask-HDR~\cite{santosSingleImageHDR2020} directly estimate an HDR image in a single step without breaking up the dynamic range. Still closely related are Single-HDR~\cite{liu2020single} and IntrinsicHDR~\cite{dilleIntrinsicHDR}, which use multi-step pipelines but go directly from linearized LDR RGB~\cite{liu2020single} or LDR Shading~\cite{dilleIntrinsicHDR} to HDR.   
They stand in contrast to DrTMO~\cite{endoDeepReverseTone2017}, Multi-Exp Gen.~\cite{le2023single}, CEVR~\cite{cevr_2023}, and DiffSyn~\cite{kim2021end}, which all create differently exposed LDR images first and combine them second via multi-exposure merging. GaSLight~\cite{gaslight}, BracketDiffusion~\cite{bemana2025bracket}, and LEDiff~\cite{wang2025lediff} follow a similar strategy, building upon fine-tuned large generative diffusion models. Our method progressively extends the dynamic range of the input to create an HDR image. We provide a numerical evaluation on two representative real-world benchmarks in Section~\ref{sec:exp:quantitative} and a detailed visual analysis of our results in Section~\ref{sec:exp:qualitative}. Further evaluation results and ablations are provided in the supplementary material.

Note that our approach operates in linear RGB space and is independent of the linearization method itself. It can be applied flexibly after dedicated linearization networks such as the one proposed by~\citet{liu2020single}, or to further extend already linear RAW images. For our numerical evaluation and all visual results, we apply an inverse gamma function~\cite{ITU-R_BT1886} as a simple baseline to linearize the images. We expect further improvements with dedicated inversion of the camera response function.

\subsection{Quantitative Evaluation}
\label{sec:exp:quantitative}
We quantitatively evaluate our approach on two HDR benchmarks. For all results, we find the best scale between the reconstructed radiance and the ground-truth based on the valid pixels in the input, and correct the CRF with a third-degree polynomial, following the protocol recommended by~\citeN{hanjiComparisonSingleImage2022}.

\paragraph{Benchmarks.}
We compare all methods on the established SI-HDR \cite{hanjiComparisonSingleImage2022} and BtP-HDR~\cite{bolduc2023beyond} benchmarks. Both sets contain diverse challenging HDR settings such as sunsets, specular reflections, and visible light sources. While SI-HDR focuses strongly on outdoor scenes, BtP-HDR fittingly extends to bright light sources in images derived from the larger Laval Indoor HDR dataset~\cite{gardner2017learning}. Both datasets are zero-shot settings for our approach and all baselines, as they have not been used in training. The only exception is GaSLight, which has been trained on BtP-HDR.

\paragraph{Metrics.}
We evaluate the results using six different metrics. PSNR and VSI are standard metrics for image quality and provide insightful results on HDR images in the perceptual uniform PU21 color space~\cite{azimi2021pu21}, which aligns them closely with human preference~\cite{hanjiComparisonSingleImage2022}. HDR-VDP3~\cite{vdp3mantiuk2023hdr}, short VDP3, is a popular HDR metric that models human perception. ColorVideoVDP~\cite{mantiuk2024colorvideovdp} (CVVDP) is closely related, with additional focus on color reproduction in HDR. FLIP~\cite{andersson2020flip} computes the reconstruction error over different exposures and is well-suited to evaluate the full dynamic range, including bright highlights. We additionally report results on PSNR-H~\cite{dilleIntrinsicHDR}, or P-H, which specifically evaluates the highlight sections of the image.

\begin{figure*}
  \includegraphics[width=\linewidth]{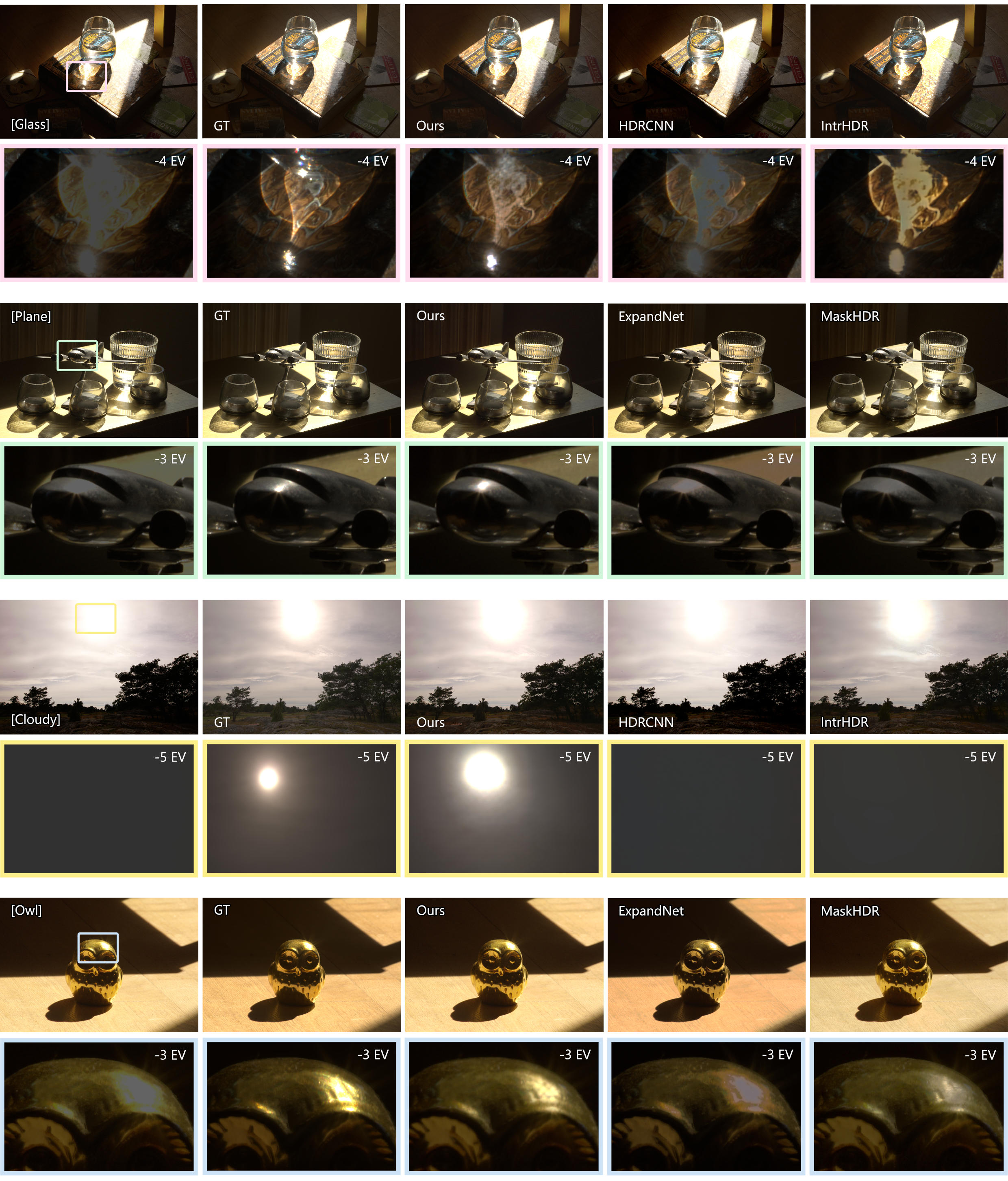}
  \caption{We show results of our method against the strongest baselines on the highlight-specific P-H metric. Our method recovers plausible highlights on SI-HDR~\cite{hanjiComparisonSingleImage2022}, where other methods saturate early or create artifacts. 
    }
  \label{fig:sihdr_results}
\end{figure*}
\begin{figure*}
  \includegraphics[width=0.98\linewidth]{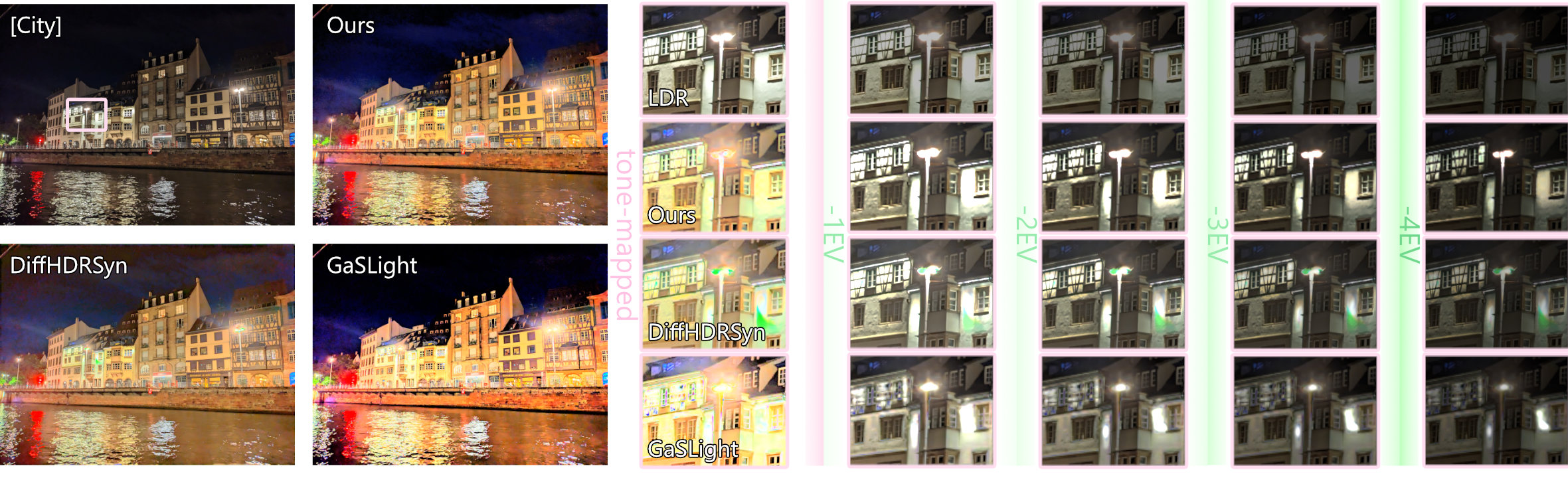}
  \includegraphics[width=0.98\linewidth]{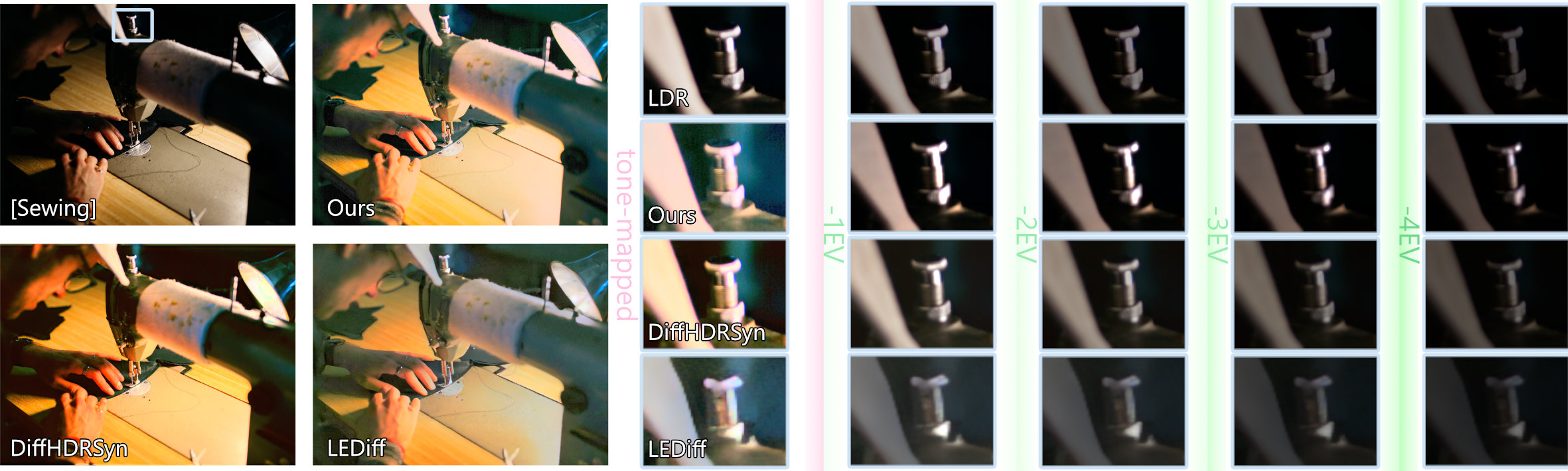}
  \caption{Compared to multi-exposure generation methods, our approach is powerful in recovering realistic luminance for bright light sources, recovers details in the highlights, and is robust against artifacts. 
  \imagecredits{Death to the Stock Photo}
  }
  \label{fig:multi_exposure_results}
\end{figure*}
\paragraph{Results.}
We show our results in~\cref{tab:all_sihdr}. Our recurrent dynamic range extension outperforms all baselines on SI-HDR with significant improvements on most metrics. The strong results on VDP3 underline the close perceptual alignment with natural scenes, supported also by the competitive results on CVVDP. The large gains on FLIP and P-H indicate that the recurrent extension enables the reconstruction of a wide dynamic range and creates realistic highlights for specular reflections and visible light sources. This is due to our exposure-invariant formulation that allows the network to learn the distribution of highlights well while maintaining robustness for lower brightnesses. With the recurrent training, the network learns meaningful radiance values over a wide dynamic range. It is robust against accumulated errors and clearly outperforms stack-reconstruction methods. This includes BracketDiffusion, LEDiff, and GaSLight, which are built on large generative diffusion networks. Our method is thus the only iterative approach that performs not only on par but better than LDR-to-HDR reconstruction baselines.

We see a similar picture emerging on BtP-HDR with more variance in the numbers. Our method performs competitively with the strongest baseline GaSLight, although the latter has been trained on the dataset. Note that neither BtP-HDR nor the original Laval Indoor HDR dataset is included in the training data for our method. This shows our generalization performance to both outdoor and indoor images and underlines once more the realistic recovery of bright light sources.

\begin{table}[t]
\centering
\caption{We compare the dynamic range gains on the highlights of the SI-HDR~\cite{hanjiComparisonSingleImage2022} benchmark.}

\resizebox{0.80\linewidth}{!}{%
\begin{tabular}{lcc}
\toprule
 Method &   avg. DR  &  max. DR  \\
\midrule
GaSLight~\tiny{\cite{gaslight}} &80,000:1 &1,049,000:1 \\
Mask-HDR~\tiny{\cite{santosSingleImageHDR2020}} &45,000:1 &1,921,000:1 \\
HDR-CNN~\tiny{\cite{eilertsenHDRImageReconstruction2017}} &50,000:1 &1,078,000:1 \\
\midrule
OURS - w/o second phase &748:1 &2,929:1 \\ 
OURS    &120,000:1 &8,778,000:1 \\
\midrule
GT      &529,000:1 &11,874,000:1 \\
\bottomrule
\end{tabular}
}
\label{tab:dr_gain}
\end{table}
\paragraph{Dynamic Range Gain.}
We evaluate the effect of our second training phase on the resulting dynamic range in~\cref{tab:dr_gain}. We show the achieved dynamic range gains on the SI-HDR~\cite{hanjiComparisonSingleImage2022} benchmark. The input is the original LDR with an encoded dynamic range of 256:1, i.e., 8 stops. Matching our formulation, we focus the evaluation on the highlight extension and mask out extended shadows in the baseline methods and the ground truth. The computed dynamic range is thus the ratio of the linearized maximum pixel value to $1/255$ as minimum. We report the average and maximum across ratios.

Without our recurrent training and HDR ground-truth data, the dynamic range extension only slightly increases the dynamic range of the input. The network has only seen RAW images in training. Used to a single extension, it is not able to reproduce the high luminance values of bright highlights. This effect also becomes visible in Figure 3 of the supplementary material, where the single-stage training saturates early and does not show further improvement for additional network executions.

Once we add the recurrent training with full HDR supervision, the performance changes significantly. Our full approach learns the formation of highlights over multiple consecutive exposures. On average, the network now performs five inference steps and more than doubles the number of stops, outperforming the strongest baselines.

Our second-stage training introduces both recurrent training and exposure to HDR ground-truth to the system, both of which help our method estimate longer-tailed distributions. We further analyze the components of our method in Section D of the supplementary document, where we show the effect of the recurrent training, the supervision with RAW data, and our two discriminators.

\begin{figure*}
\footnotesize
  \resizebox{\linewidth}{!}{
    \begin{tabular}
    {  
        p{0.2\linewidth}
        p{0.2\linewidth}
        p{0.2\linewidth}
        p{0.2\linewidth}
        p{0.2\linewidth}
    }
\centering Input image (LDR) & \centering Single-HDR~\cite{liu2020single} & \centering Mask-HDR~\cite{santosSingleImageHDR2020} & \centering HDRUNet~\cite{chen2021hdrunet} & \centering Ours
    \end{tabular}
    }
  \includegraphics[width=0.98\linewidth]{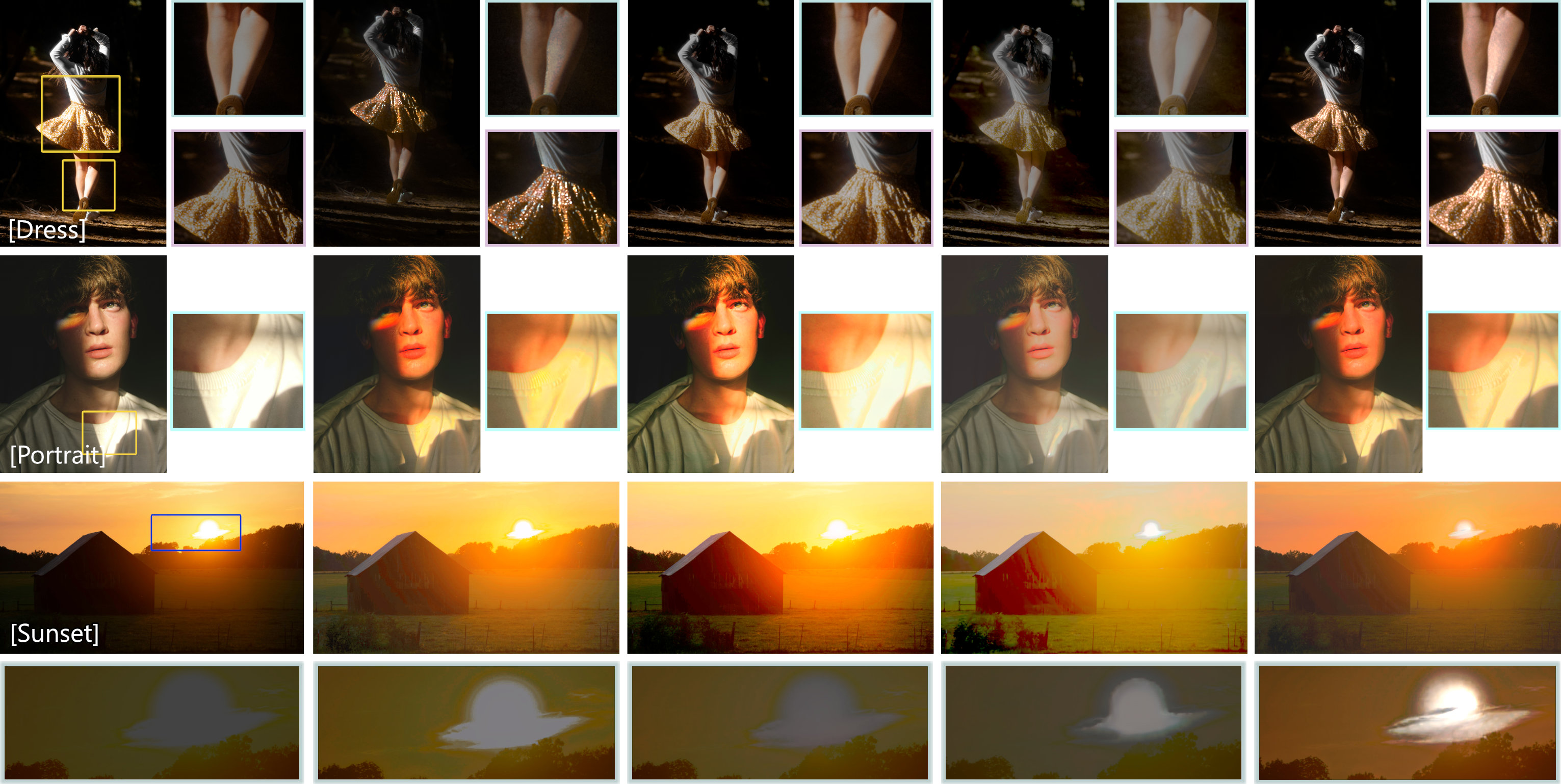}
  \caption{Our method faithfully recovers clipped areas in images in the wild with colors that match the environment. The full images are tone-mapped.
  \newline \imagecredits{
  \myhref[darkgray]{https://unsplash.com/photos/a-woman-in-a-yellow-skirt-is-walking-in-the-dark-02gX51Gy1f0}{@$\text{Matthew Ball}$}, \myhref[darkgray]{https://unsplash.com/photos/man-in-white-crew-neck-shirt-0rYi_P711Hk}{@$\text{gryffyn m}$},  \myhref[darkgray]{https://unsplash.com/photos/the-sun-is-setting-behind-a-barn-in-a-field-ktlpp1oEB5c}{@$\text{Elmer Cañas}$}
  }
  } 
  \label{fig:itw_results} 
\end{figure*}

\subsection{Qualitative Evaluation}
\label{sec:exp:qualitative}
We support our numerical findings with visual comparisons of examples from the SI-HDR~\cite{hanjiComparisonSingleImage2022} benchmark and images in the wild in Figures~\ref{fig:sihdr_results}, ~\ref{fig:multi_exposure_results}, and~\ref{fig:itw_results}.
Our method performs strongly in recovering the highlights even in extreme cases, and simultaneously inpaints clipped areas convincingly in a wide range of scenarios.

\paragraph{Highlight Reconstruction.}
Recovering the accurate brightness of highlights is one of the main challenges for HDR reconstruction. Their extreme values create the long-tailed distribution of the HDR image and force direct LDR-to-HDR reconstruction methods to compress the luminance range during training. We show the results of our method on challenging highlights in the SI-HDR benchmark in~\cref{fig:sihdr_results}. For each example, we show the ground-truth and the results from HDRCNN, MaskHDR, IntrinsicHDR, and ExpandNet, all of which demonstrate competitive performance in quantitative baselines (\cref{sec:exp:quantitative}).

In the [\texttt{Glass}] example, the LDR input shows clipping on the book cover where it is hit by the sunlight reflected from the glass. The ground-truth reveals a bright specular spot in the lower part of the clipped area, which remains clipped even when the exposure is lowered by four stops. In the LDR input, the large brightness difference between the specular highlight and the book cover is lost due to clipping. For this reason, both baseline methods equally increase the brightness of the full clipped area and struggle to reproduce the highlight spot. Being trained on a compressed dynamic range, they saturate early, resulting in flat gray regions on the book cover. In contrast, our approach faithfully matches the brightness of the specular highlight. It also inpaints parts of the lost texture of the book cover, while recovering the challenging contrast between the book cover and the specular highlight present in the ground-truth. 

We observe a similar behavior for the [\texttt{Plane}] and the [\texttt{Owl}] scenes. For the [\texttt{Plane}], ExpandNet and MaskHDR both leave the clipped highlight unchanged. For the [\texttt{Owl}], MaskHDR faithfully recovers the golden metallic appearance of the owl, but does not reproduce the brightness of the highlight. ExpandNet creates color artifacts and barely increases the dynamic range of the input. In both cases, our method recovers the strong direct reflection of the nose of the plane and the golden head of the owl and extends the dynamic range beyond four exposure values, in accordance with the ground-truth. 

[\texttt{Cloudy}] poses a challenge, as the sun is visible through the cloud cover of the sky. The bright spot of the sun in the ground-truth is in contrast to the overcast sky in the surroundings. As a result, both HDRCNN and IntrinsicHDR reconstruct a cloudy sky with low dynamic range, while our method creates a realistic sun disc with corresponding brightness. Note that our generated dynamic range exceeds the number of recurrent steps used during training. 

The examples show the strength of our extension formulation in being adaptive to the content of the scene rather than being limited to the specific number of exposure brackets used during training. Our formulation leaves the luminance uncompressed, and the method learns to recover accurate bright highlights even in these challenging scenarios.

\paragraph{Multi-Exposure Consistency.}
Multi-exposure reconstruction approaches require multiple inference steps of the same network. This is prone to accumulating prediction errors and creating artifacts in the estimated result. We compare our results with the recent exposure-stack generation methods DiffHDRSyn, LEDiff, and GaSLight over a wide dynamic range in Figure~\ref{fig:multi_exposure_results}. The last two baselines are both based on pretrained generative diffusion models that have been fine-tuned on HDR reconstruction. 

For the [\texttt{City}], DiffHDRSyn and GaSLight both fail to preserve the brightness of the light source. DiffHDRSyn additionally creates strong color artifacts in the clipped areas, while GaSLight exaggerates the brightness of the wall, giving it the appearance of an active light source. Both cases present examples of artifacts caused by iterative inference. In the [\texttt{Sewing}] image, DiffHDRSyn and LEDiff only slightly extend the brightness of the pin, resulting in a more diffuse, non-metallic appearance. Notably, the generative LEDiff incorrectly modifies the shape of the pin, which is a common weakness of diffusion-based methods. 

In comparison, our method maintains the strong brightness of the light source in the [\texttt{City}] example and reduces the brightness of the diffuse wall realistically while recovering the color. This shows the importance of our recurrent training, which stabilizes the dynamic range extension over multiple inference steps and reduces accumulated artifacts such as those of the other two methods.
For this reason, our method reproduces the metallic appearance of the pin in the [\texttt{Sewing}] example, preserves its shape, and concentrates the intensity of the specular highlight realistically on a single point. 

Our method maintains accurate brightness levels for light sources, bright surfaces, and reflections over multiple exposure steps. We show more examples in our supplementary material.

\paragraph{Detail Recovery.}
We compare the ability to recover clipped details on images in the wild in~\cref{fig:itw_results}. Generalization to diverse imagery and realistic inpainting of details is another challenge for HDR reconstruction methods, since learning a strong prior on natural images is limited by the scarcity of HDR training data. We compare against LDR-to-HDR reconstruction methods SingleHDR, MaskHDR, and HDRUnet, and show additional results against IntrinsicHDR, DiffHDRSyn, and LEDiff in the supplementary.

The clipped LDR input of the [$\texttt{Dress}$] requires recovery of both skin color and the pattern on the skirt. Being trained on HDR images alone, SingleHDR produces extremely bright, unnatural spots for the skirt. MaskHDR and HDRUnet, on the other hand, both leave the skin clipped and fail to recover natural color. For the [\texttt{Portrait}], SingleHDR and HDRUnet both fail to accurately inpaint the shirt and produce artifacts. MaskHDR leaves the clipped area completely white, creating a clear boundary. Our method, in contrast, performs well on the [\texttt{Dress}], recovering the color of the skirt and the skin tone. The estimated values match the LDR input without being compressed. We observe similar performance for the [\texttt{Portrait}]: Our method seamlessly fills the clipped areas with matching colors.  
We attribute the behavior to our training on diverse RAW collections that enables our network to learn a prior over natural images. This prior is maintained through the dynamic range extension, so that our method can add details even in bright highlight regions. This becomes visible for the Church example in our teaser~\cref{fig:teaser}, and for the [\texttt{Sunset}] example in Figure~\ref{fig:itw_results}, where it is the only method reproducing the shape of the cloud in front of the sun.
\begin{figure}
  \includegraphics[width=\linewidth]{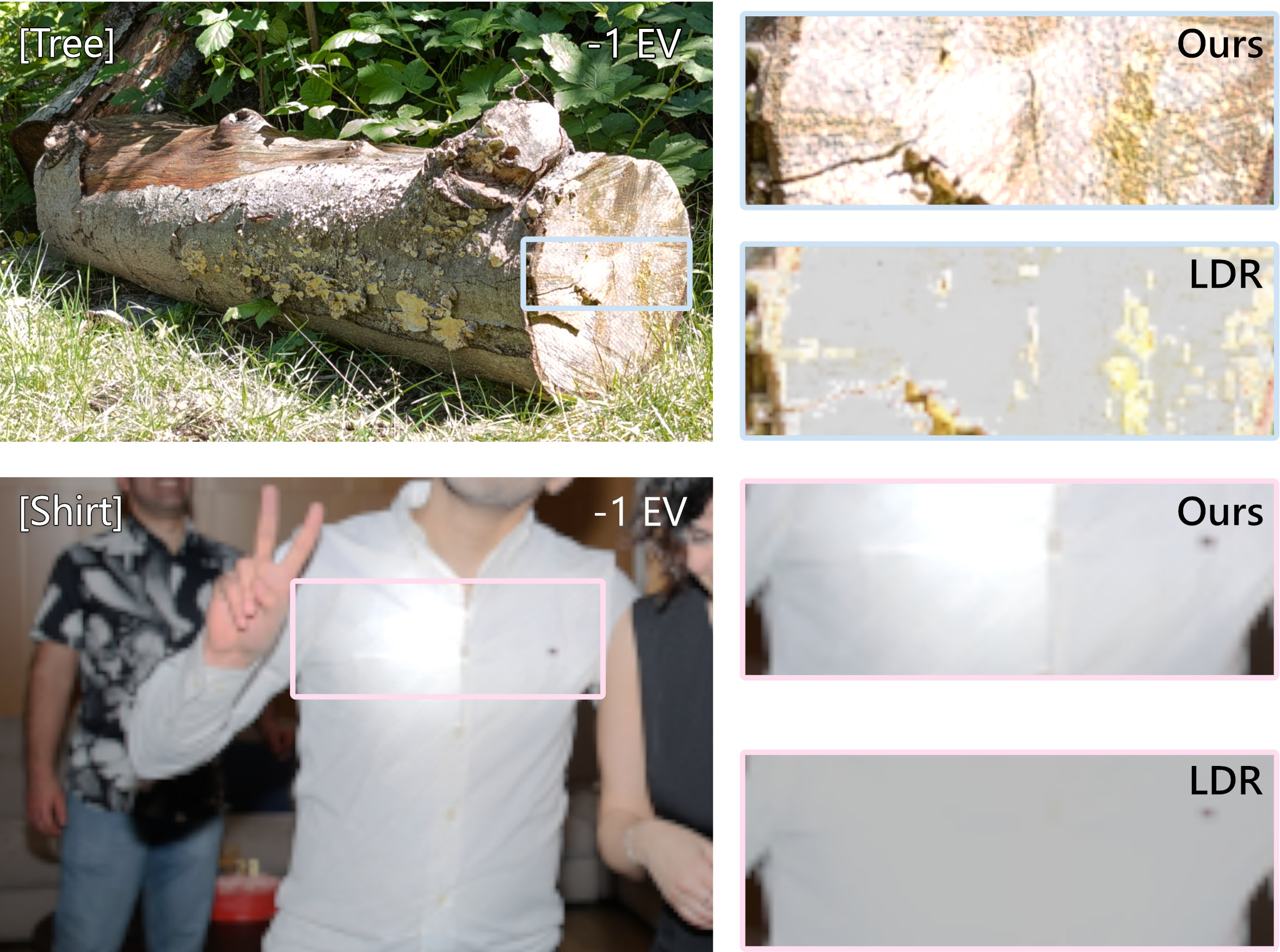}
  \caption{
     Limitations of our method. While the network creates local structure and details, it results in a repetitive, unrealistic pattern for the tree. In some cases, our method overestimates the brightness of the clipped area, creating a bright spot on the shirt.
  }
  \label{fig:limitations}
\end{figure}

\section{Conclusion And Limitations}
\label{sec:conclusion}
We have presented a recurrent dynamic range extension approach that progressively recovers the highlights of an HDR image. Our method generalizes to diverse in-the-wild scenarios by faithfully recovering skin tones, specular reflections, and light sources. The generalization performance comes from our scale-invariant formulation that enables training on rich RAW photo collections. Our recurrent training setup opens the extension to arbitrary dynamic ranges without restriction during training. We demonstrate the performance in an extensive visual and numerical analysis.

We show limitations of our approach in \cref{fig:limitations}. Stemming from the distribution of RAW image collections, our approach assumes casual photography conditions with reasonably exposed midtones. Under these circumstances, our extension network generates structure matching to the local neighbourhood. For the severe overexposure on the [\texttt{Tree}] trunk with larger clipped areas of the complex structure of the bark, however, its generative capabilities are limited and can lead to repetitive patterns. Similarly, our approach can overestimate the original brightness of affected areas. In the case of the [\texttt{Shirt}], the LDR image contains a very large area of all-white pixels, which does not provide sufficient semantic information for our approach to reason about the context. Our method correctly extends the wrinkles of the cloth, but creates an unrealistic brighter light spot in the center.

Further, our method addresses one part of the larger HDR reconstruction task. Recovering the full radiance of a scene additionally involves the removal of noise and quantization artifacts in the shadows, as well as the inversion of the non-linear CRF. These effects have different characteristics from clipping and pose interesting targets for future work.

\begin{acks}
We thank Alireza Moazeni and Ke Li for our discussions in the early stages of this work. We acknowledge the support of the Natural Sciences and Engineering Research Council of Canada (NSERC), [RGPIN-2020-05375].
\end{acks}

\bibliographystyle{ACM-Reference-Format}
\bibliography{main}
\end{document}